\pdfoutput=1
\documentclass[12pt,a4paper]{article}

\usepackage{ifthen} 
\newboolean{pdflatex}
\setboolean{pdflatex}{true} 

\newboolean{articletitles}
\setboolean{articletitles}{true} 

\newboolean{uprightparticles}
\setboolean{uprightparticles}{false} 

\def\paperauthors{LHCb collaboration} 
\def\paperasciititle{Study of charmonium production via the decay to ppbar at sqs = 13 TeV} 
\def\papertitle{Study of charmonium production via the decay to $p\bar{p}$ at $\sqrt{s} = 13\,\rm{TeV}$} 
\def\paperkeywords{{High Energy Physics}, {LHCb}, {QCD}, {Quarkonium}, {Particle and resonance production}, {Branching fraction}} 
\def\papercopyright{\the\year\ CERN for the benefit of the LHCb collaboration} 
\def\paperlicence{CC BY 4.0 licence}
\def\paperlicenceurl{https://creativecommons.org/licenses/by/4.0/}

\usepackage[top=1in, bottom=1.25in, left=1in, right=1in]{geometry}

\usepackage{microtype}
\usepackage{lineno}  
\usepackage{xspace} 
\usepackage{caption} 

\usepackage{graphicx}  
\usepackage{color}
\usepackage{colortbl}
\graphicspath{{./figs/}} 

\usepackage{amsmath} 
\usepackage{amssymb}
\usepackage{amsfonts}
\usepackage{upgreek} 

\newcommand*\patchAmsMathEnvironmentForLineno[1]{%
\expandafter\let\csname old#1\expandafter\endcsname\csname #1\endcsname
\expandafter\let\csname oldend#1\expandafter\endcsname\csname
end#1\endcsname
 \renewenvironment{#1}%
   {\linenomath\csname old#1\endcsname}%
   {\csname oldend#1\endcsname\endlinenomath}%
}
\newcommand*\patchBothAmsMathEnvironmentsForLineno[1]{%
  \patchAmsMathEnvironmentForLineno{#1}%
  \patchAmsMathEnvironmentForLineno{#1*}%
}
\AtBeginDocument{%
\patchBothAmsMathEnvironmentsForLineno{equation}%
\patchBothAmsMathEnvironmentsForLineno{align}%
\patchBothAmsMathEnvironmentsForLineno{flalign}%
\patchBothAmsMathEnvironmentsForLineno{alignat}%
\patchBothAmsMathEnvironmentsForLineno{gather}%
\patchBothAmsMathEnvironmentsForLineno{multline}%
\patchBothAmsMathEnvironmentsForLineno{eqnarray}%
}

\usepackage{hyperxmp}

\usepackage[pdftex,
            pdfauthor={\paperauthors},
            pdftitle={\paperasciititle},
            pdfkeywords={\paperkeywords},
            pdfcopyright={Copyright (C) \papercopyright},
            pdflicenseurl={\paperlicenceurl}]{hyperref}

\usepackage[colorinlistoftodos,textsize=scriptsize]{todonotes}

\usepackage[bottom,flushmargin,hang,multiple]{footmisc}

\usepackage[all]{hypcap} 

\usepackage{xspace} 
\usepackage{upgreek}

\def\lhcb   {\mbox{LHCb}\xspace}

\def\MagUp {\mbox{\em Mag\kern -0.05em Up}\xspace}

\ifthenelse{\boolean{uprightparticles}}%
{

 \def\Peta        {\ensuremath{\upeta}\xspace}

 \def\Ppi         {\ensuremath{\uppi}\xspace}

 \def\Pchi        {\ensuremath{\upchi}\xspace}                 
 \def\Ppsi        {\ensuremath{\uppsi}\xspace}

 \def\PDelta      {\ensuremath{\Delta}\xspace}                 
 \def\PXi         {\ensuremath{\Xi}\xspace}                 
 \def\PLambda     {\ensuremath{\Lambda}\xspace}                 
 \def\PSigma      {\ensuremath{\Sigma}\xspace}                 
 \def\POmega      {\ensuremath{\Omega}\xspace}                 
 \def\PUpsilon    {\ensuremath{\Upsilon}\xspace}
 \let\oldPi\Pi
 \def\PPi         {\ensuremath{\oldPi}\xspace}

 \def\PB      {\ensuremath{\mathrm{B}}\xspace}                 
                  
 \def\PD      {\ensuremath{\mathrm{D}}\xspace}

 \def\PJ      {\ensuremath{\mathrm{J}}\xspace}                 
 \def\PK      {\ensuremath{\mathrm{K}}\xspace}

 \def\Pb      {\ensuremath{\mathrm{b}}\xspace}                 
 \def\Pc      {\ensuremath{\mathrm{c}}\xspace}

 \def\Ph      {\ensuremath{\mathrm{h}}\xspace}                 
 \def\Pi      {\ensuremath{\mathrm{i}}\xspace}

 \def\Pp      {\ensuremath{\mathrm{p}}\xspace}

 \def\Ps      {\ensuremath{\mathrm{s}}\xspace}

 \def\Py      {\ensuremath{\mathrm{y}}\xspace}                 
                  
 \def\thebaroffset{0.0em}
}
{

 \def\Peta        {\ensuremath{\eta}\xspace}

 \def\Ppi         {\ensuremath{\pi}\xspace}

 \def\Pchi        {\ensuremath{\chi}\xspace}                 
 \def\Ppsi        {\ensuremath{\psi}\xspace}                 
                  
 \mathchardef\PDelta="7101
 \mathchardef\PXi="7104
 \mathchardef\PLambda="7103
 \mathchardef\PSigma="7106
 \mathchardef\POmega="710A
 \mathchardef\PUpsilon="7107
 \mathchardef\PPi="7105
                  
 \def\PB      {\ensuremath{B}\xspace}                 
                  
 \def\PD      {\ensuremath{D}\xspace}

 \def\PJ      {\ensuremath{J}\xspace}                 
 \def\PK      {\ensuremath{K}\xspace}

 \def\Pb      {\ensuremath{b}\xspace}                 
 \def\Pc      {\ensuremath{c}\xspace}

 \def\Ph      {\ensuremath{h}\xspace}                 
 \def\Pi      {\ensuremath{i}\xspace}

 \def\Pp      {\ensuremath{p}\xspace}

 \def\Ps      {\ensuremath{s}\xspace}

 \def\Py      {\ensuremath{y}\xspace}                 
 
 \def\thebaroffset{0.18em}
}
\newcommand{\offsetoverline}[2][\thebaroffset]{\kern #1\overline{\kern -#1 #2}}%

\makeatletter
\ifcase \@ptsize \relax
  \newcommand{\miniscule}{\@setfontsize\miniscule{4}{5}}
\or
  \newcommand{\miniscule}{\@setfontsize\miniscule{5}{6}}
\or
  \newcommand{\miniscule}{\@setfontsize\miniscule{5}{6}}
\fi
\makeatother

\DeclareRobustCommand{\optbar}[1]{\shortstack{{\miniscule (\rule[.5ex]{1.25em}{.18mm})}
  \\ [-.7ex] $#1$}}

\def\squark    {{\ensuremath{\Ps}}\xspace}

\def\cquark    {{\ensuremath{\Pc}}\xspace}
\def\cquarkbar {{\ensuremath{\overline \cquark}}\xspace}
\def\ccbar     {{\ensuremath{\cquark\cquarkbar}}\xspace}
\def\bquark    {{\ensuremath{\Pb}}\xspace}

\def\pion   {{\ensuremath{\Ppi}}\xspace}
\def\piz    {{\ensuremath{\pion^0}}\xspace}

\def\KorKbar {\kern \thebaroffset\optbar{\kern -\thebaroffset \PK}{}\xspace}

\def\D       {{\ensuremath{\PD}}\xspace}

\def\DorDbar {\kern \thebaroffset\optbar{\kern -\thebaroffset \PD}\xspace}

\def\Dp      {{\ensuremath{\D^+}}\xspace}
\def\Dm      {{\ensuremath{\D^-}}\xspace}

\def\DpDm    {\ensuremath{\Dp {\kern -0.16em \Dm}}\xspace}

\def\B       {{\ensuremath{\PB}}\xspace}

\def\BorBbar {\kern \thebaroffset\optbar{\kern -\thebaroffset \PB}\xspace}

\def\Bd      {{\ensuremath{\B^0}}\xspace}

\def\BdorBdbar {\kern \thebaroffset\optbar{\kern -\thebaroffset \Bd}\xspace}

\def\Bs      {{\ensuremath{\B^0_\squark}}\xspace}

\def\BsorBsbar {\kern \thebaroffset\optbar{\kern -\thebaroffset \Bs}\xspace}

\def\jpsi     {{\ensuremath{{\PJ\mskip -3mu/\mskip -2mu\Ppsi}}}\xspace}
\def\psitwos  {{\ensuremath{\Ppsi{(2S)}}}\xspace}

\def\etac     {{\ensuremath{\Peta_\cquark}}\xspace}

\def\chiczero {{\ensuremath{\Pchi_{\cquark 0}}}\xspace}
\def\chicone  {{\ensuremath{\Pchi_{\cquark 1}}}\xspace}
\def\chictwo  {{\ensuremath{\Pchi_{\cquark 2}}}\xspace}
\def\chicJ    {{\ensuremath{\Pchi_{\cquark J}}}\xspace}

\def\Y#1S{\ensuremath{\PUpsilon{(#1S)}}\xspace}

\def\proton      {{\ensuremath{\Pp}}\xspace}
\def\antiproton  {{\ensuremath{\overline \proton}}\xspace}

\def\LorLbar     {\kern \thebaroffset\optbar{\kern -\thebaroffset \PLambda}\xspace}

\def\BF         {{\ensuremath{\mathcal{B}}}\xspace}
\def\BR         {\BF}

\newcommand{\decay}[2]{\ensuremath{#1\!\to #2}\xspace} 

\def\to                 {\ensuremath{\rightarrow}\xspace}

\def\AT#1     {\ensuremath{A_{\mathrm{T}}^{#1}}\xspace}           

\def\C#1      {\ensuremath{\mathcal{C}_{#1}}\xspace}                       
\def\Cp#1     {\ensuremath{\mathcal{C}_{#1}^{'}}\xspace}                    
\def\Ceff#1   {\ensuremath{\mathcal{C}_{#1}^{\mathrm{(eff)}}}\xspace}        
\def\Cpeff#1  {\ensuremath{\mathcal{C}_{#1}^{'\mathrm{(eff)}}}\xspace}       
\def\Ope#1    {\ensuremath{\mathcal{O}_{#1}}\xspace}                       
\def\Opep#1   {\ensuremath{\mathcal{O}_{#1}^{'}}\xspace}                    

\newcommand{\nospaceunit}[1]{\ensuremath{\text{#1}}}       
\newcommand{\aunit}[1]{\ensuremath{\text{\,#1}}}       

\newcommand{\tev}{\aunit{Te\kern -0.1em V}\xspace}
\newcommand{\gev}{\aunit{Ge\kern -0.1em V}\xspace}
\newcommand{\mev}{\aunit{Me\kern -0.1em V}\xspace}
\newcommand{\kev}{\aunit{ke\kern -0.1em V}\xspace}
\newcommand{\ev}{\aunit{e\kern -0.1em V}\xspace}
 
\newcommand{\mevc}{\ensuremath{\aunit{Me\kern -0.1em V\!/}c}\xspace}
\newcommand{\gevc}{\ensuremath{\aunit{Ge\kern -0.1em V\!/}c}\xspace}
\newcommand{\mevcc}{\ensuremath{\aunit{Me\kern -0.1em V\!/}c^2}\xspace}
\newcommand{\gevcc}{\ensuremath{\aunit{Ge\kern -0.1em V\!/}c^2}\xspace}

\def\mum  {\ensuremath{\,\upmu\nospaceunit{m}}\xspace}

\def\nb {\aunit{nb}\xspace}

\def\fb   {\ensuremath{\aunit{fb}}\xspace}
\def\invfb   {\ensuremath{\fb^{-1}}\xspace}

\def\fs   {\aunit{fs}}

\def\gsim{{~\raise.15em\hbox{$>$}\kern-.85em
          \lower.35em\hbox{$\sim$}~}\xspace}
\def\lsim{{~\raise.15em\hbox{$<$}\kern-.85em
          \lower.35em\hbox{$\sim$}~}\xspace}

\def\sqs   {\ensuremath{\protect\sqrt{s}}\xspace}

\def\pt         {\ensuremath{p_{\mathrm{T}}}\xspace}

\def\ptot       {\ensuremath{p}\xspace}

\def\tell1  {TELL1\xspace}
\def\ukl1   {UKL1\xspace}

\newcommand{\eg}{\mbox{\itshape e.g.}\xspace}

\newcommand{\lhcborcid}[1]{\href{https://orcid.org/#1}{\hspace*{0.1em}\raisebox{-0.45ex}{\includegraphics[width=1em]{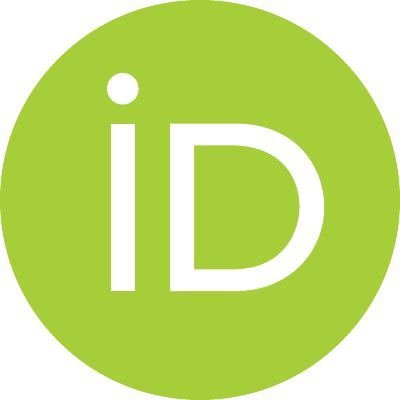}}}}

\usepackage{cite} 
\usepackage{mciteplus}

\usepackage{longtable} 
\usepackage{rotating} 
\usepackage{multirow} 
\usepackage{booktabs} 
\def\etacones     {{\ensuremath{\Peta_\cquark{(1S)}}}\xspace}
\def\etactwos     {{\ensuremath{\Peta_\cquark{(2S)}}}\xspace}
\def\hc           {{\ensuremath{\Ph_\cquark{(1P)}}}\xspace}
\def\ppbar     {\ensuremath{\proton\antiproton}\xspace}

\def\y     {\ensuremath{\Py}\xspace}

\def\JpsiToPpbar        {\ensuremath{\jpsi\to\ppbar}\xspace}
\def\EtacToPpbar        {\ensuremath{\etac\to\ppbar}\xspace}

\def\JpsiToPpbarPiz     {\ensuremath{\jpsi\to\ppbar\piz}\xspace}

\begin{document}

\renewcommand{\thefootnote}{\fnsymbol{footnote}}
\setcounter{footnote}{1}


\begin{titlepage}
\pagenumbering{roman}

\vspace*{-1.5cm}
\centerline{\large EUROPEAN ORGANIZATION FOR NUCLEAR RESEARCH (CERN)}
\vspace*{1.5cm}
\noindent
\begin{tabular*}{\linewidth}{lc@{\extracolsep{\fill}}r@{\extracolsep{0pt}}}
\ifthenelse{\boolean{pdflatex}}
{\vspace*{-1.5cm}\mbox{\!\!\!\includegraphics[width=.14\textwidth]{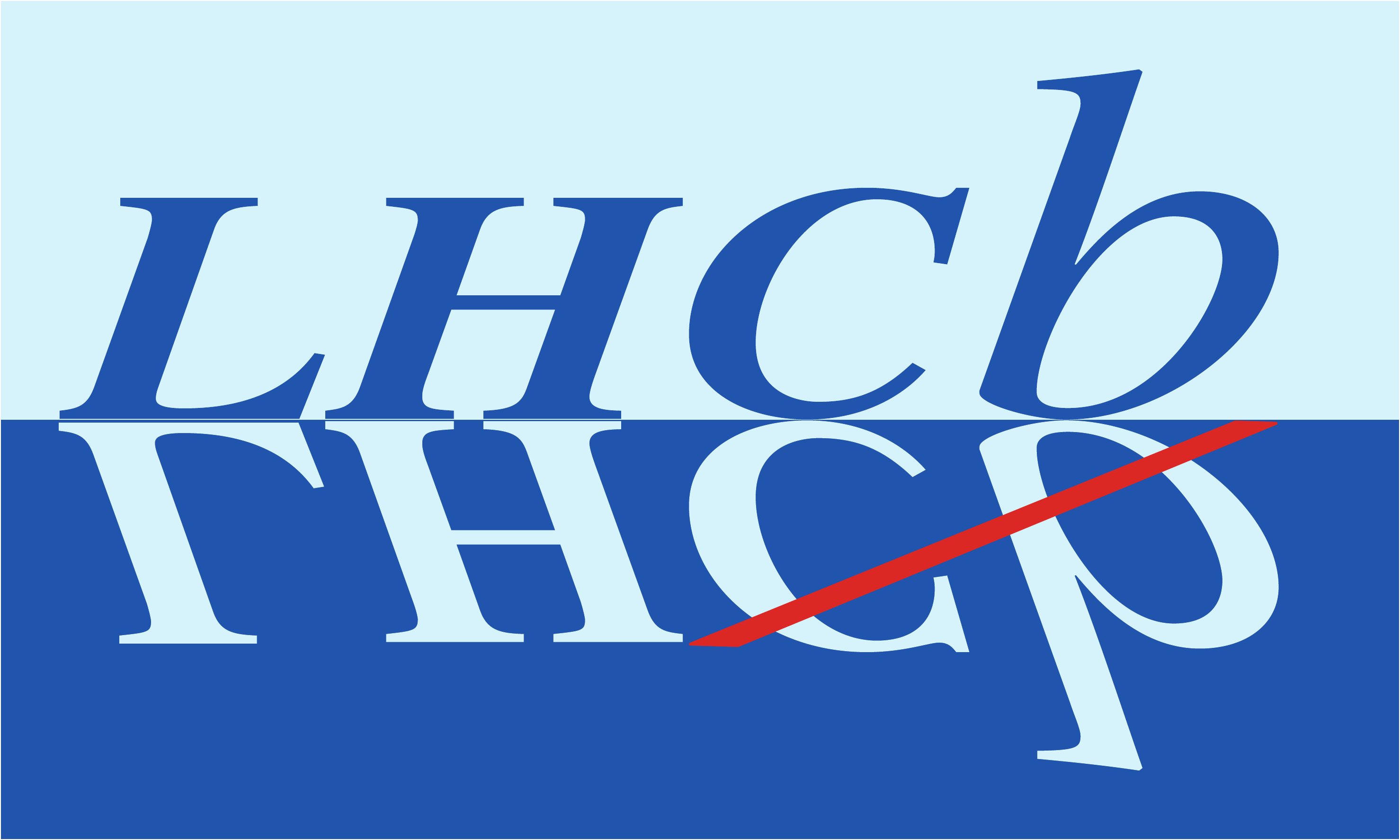}} & &}%
{\vspace*{-1.2cm}\mbox{\!\!\!\includegraphics[width=.12\textwidth]{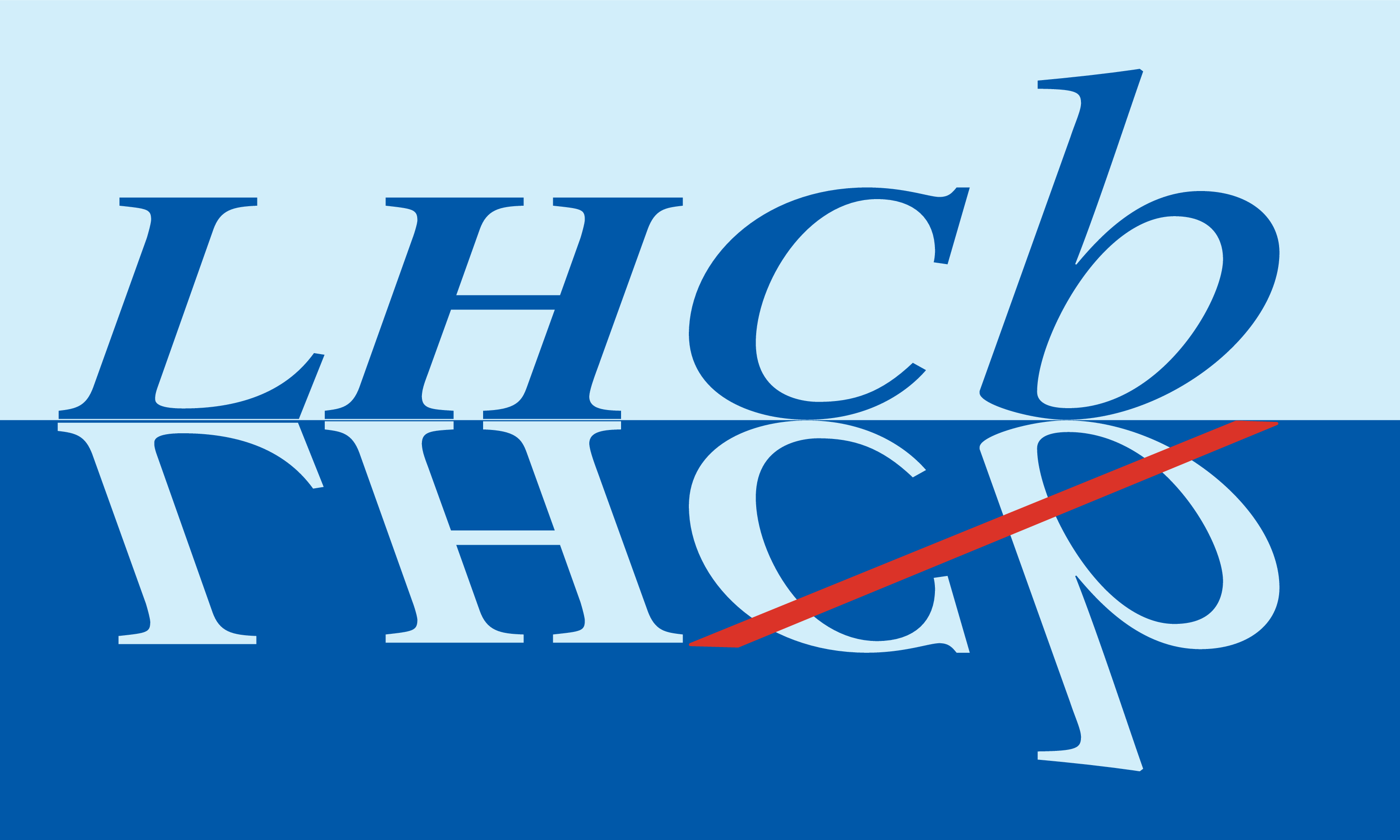}} & &}%
\\
 & & CERN-EP-2024-165 \\  
 & & LHCb-PAPER-2024-004 \\  
 & & January 27, 2025 \\ 
 & & \\
\end{tabular*}

\vspace*{4.0cm}

{\normalfont\bfseries\boldmath\huge
\begin{center}
  \papertitle 
\end{center}
}

\vspace*{2.0cm}

\begin{center}
\paperauthors\footnote{Authors are listed at the end of this paper.}
\end{center}

\vspace{\fill}

\begin{abstract}
  \noindent
  Charmonium production cross-section in proton-proton collisions is measured at the centre-of-mass energy $\sqrt{s}=13\,\rm{TeV}$ using decays to $p\bar{p}$ final state.
  The study is performed using a data sample corresponding to an integrated luminosity of $2.2\,\rm{fb}^{-1}$ collected in 2018 with the $\mbox{LHCb}$ detector.
  The production cross-section of the $\eta_c$ meson is measured in a rapidity range of $2.0 < y < 4.0$ and in a transverse momentum range of \mbox{$5.0 < p_{\rm{T}} < 20.0\,\rm{GeV\!/\!\it{c}}$}, which is extended compared with previous $\mbox{LHCb}$ analyses. 
  The differential cross-section is measured in bins of $p_{\rm{T}}$ and, for the first time, of $y$. 
  Upper limits, at 90\% and 95\% confidence levels, on the $\eta_c(2S)$ and $h_c(1P)$ prompt production cross-sections are determined for the first time. 
\end{abstract}

\vspace*{2.0cm}

\begin{center}
  Published in 
  \href{https://doi.org/10.1140/epjc/s10052-024-13478-y}{Eur.~Phys.~J.~C84 (2024) 1274}
\end{center}

\vspace{\fill}

{\footnotesize 
\centerline{\copyright~\papercopyright. \href{\paperlicenceurl}{\paperlicence}.}}
\vspace*{2mm}

\end{titlepage}


\newpage
\setcounter{page}{2}
\mbox{~}
%
%
%
%


\renewcommand{\thefootnote}{\arabic{footnote}}
\setcounter{footnote}{0}

\cleardoublepage


\pagestyle{plain} 
\setcounter{page}{1}
\pagenumbering{arabic}


\section{Introduction}
\label{sec:Introduction}
 
\par
In proton-proton (\proton\proton) collision events at the LHC, charmonium can originate from hadroproduction in a primary \proton\proton interaction vertex (PV)  (referred to as ``prompt production" hereafter) or from \bquark-hadron decays in a vertex displaced from the PV by the flight distance of the \bquark hadron (referred to as ``from b-hadron"). 
In both cases, charmonium may also originate from decays of intermediate excited charmonium states (``feed-down"), 
whose contributions can only be subtracted when the production cross-section of the excited state and the probability of transition between the two charmonium states are known. 
\par
Despite a substantial theoretical effort, no consistent picture of the charmonium production mechanism is available~\cite{Lansberg:2019adr}.
Of the available approaches, nonrelativistic quantum chromodynamics (NRQCD) is the most successful in describing a wide range of measurements.
It relies on three main assumptions.
First, it assumes a factorisation between the perturbative process that creates a \cquark quark and \cquarkbar antiquark, and the nonperturbative hadronisation of this pair into a charmonium meson. 
Second, it assumes the universality of the nonperturbative long-distance matrix elements (LDMEs), which considers, \eg, charmonium hadroproduction and production in \bquark-hadron decays.
Finally, the assumption of heavy-quark spin symmetry (HQSS) provides relations between LDMEs for the charmonium states with the same orbital and radial quantum numbers, \eg, the \jpsi and \etacones mesons.\footnote{The \etacones state is referred to as $\etac$ throughout the paper.} 

The NRQCD calculations at next-to-leading order (NLO) successfully describe the observed production and polarisation of the \jpsi and \psitwos states, but only in a limited range of charmonium transverse momentum \pt. 
A consistent description of the \jpsi hadro- and photo-production and polarisation, together with the production of the \etac meson, remains challenging~\cite{Butenschoen:2014dra}. 
The first measurement of the \etac meson production by the LHCb collaboration \cite{LHCb-PAPER-2014-029} showed that the colour singlet (CS) contribution saturates the observed \pt-differential cross-section in the studied \pt region. 
This result led to a major revisiting of the theoretical framework and yielded new approaches capable of simultaneously describing the \jpsi and \etac production and \jpsi polarisation, albeit still within a limited \pt range~\cite{Han:2014jya}. 
The recent \lhcb measurement of the \etac production~\cite{LHCb-PAPER-2019-024} further constrains the theory and indicates a possible colour octet (CO) contribution at high \pt.
Meanwhile, the theoretical prediction of Ref.\cite{Lansberg:2020ejc} demonstrated the unphysical behaviour of the theoretical description at low \pt leading to negative total \etac production cross-section. 
The latter motivates a new measurement of \etac production at low \pt. 
\par
In order to confront the measured production cross-sections of the \etac and \jpsi mesons with theory, knowledge of the feed-down from excited states is needed. 
Instead, the authors of Ref.\cite{Lansberg:2017ozx} proposed to measure the production of the first excited states, \etactwos and \psitwos mesons, free from feed-down contributions. 
The paper suggests the same relations between LDMEs as for the \etac and \jpsi states. 
In the same work, the authors predicted the \etactwos hadroproduction cross-section in the LHCb acceptance at 13\tev, based on three different sets of LDMEs from Refs.\cite{Shao:2014yta,Gong:2012ug,Bodwin:2015iua}.
\par
While $J^{PC}=1^{- -}$ charmonium states,  \jpsi and \psitwos mesons, are precisely studied thanks to their reconstruction in the experimentally clean dimuon decays, a simultaneous reconstruction of all known charmonium states is only possible via their decays to hadrons. 
In this paper, charmonium decays to \ppbar are used, following previous studies done by the \lhcb experiment~\cite{LHCb-PAPER-2014-029,LHCb-PAPER-2019-024}. 
This decay is available for all charmonium states~\cite{PDG2022} and thus allows studying them simultaneously.
The analysis covers prompt charmonium production and charmonium production in \bquark-hadron decays of \etac, \etactwos, \chicJ and \hc states,
using data collected by the \lhcb experiment in 2018 corresponding to an integrated luminosity of 2.2\invfb.
The differential production cross-section of \etac is measured in four bins of rapidity, in the $2.0<\y<4.0$ range, and in six \pt bins, in the $5.0<\pt<20.0 \gevc$ range, which extends that of Ref.\cite{LHCb-PAPER-2019-024}.
Upper limits are set for the prompt \etactwos and \hc production cross-sections.
The results also include the branching fractions of the \chicJ production in inclusive \bquark-hadron decays.
\par
With its precise vertex reconstruction, powerful particle identification (PID) and flexible trigger, the \lhcb experiment possesses excellent capabilities for charmonium production studies \cite{LHCb-PAPER-2014-029,LHCb-PAPER-2019-024,LHCb-PAPER-2012-039,LHCb-PAPER-2011-003,LHCb-PAPER-2013-016,LHCb-PAPER-2015-037,LHCb-PAPER-2011-045,LHCb-PAPER-2018-049,LHCb-PAPER-2011-019,LHCb-PAPER-2013-028}. 
This suppresses a large combinatorial background contribution originating from the \proton\proton interaction. 
The present study exploits the 2018 data sample with a dedicated software trigger selection of \ppbar combinations compatible with charmonium decays, in an extended range of invariant mass, transverse momentum and with reduced PID requirements compared to previous \lhcb measurements~\cite{LHCb-PAPER-2014-029,LHCb-PAPER-2019-024}. 
\par

\section{The LHCb detector and data sample}
\label{sec:Detector}
\par
The \lhcb detector~\cite{LHCb-DP-2008-001,LHCb-DP-2014-002} is a single-arm forward spectrometer covering the \mbox{pseudorapidity} range $2<\eta <5$,
designed for the study of particles containing \bquark or \cquark quarks. 
The detector includes a high-precision tracking system
consisting of a silicon-strip vertex detector surrounding the $pp$ interaction region, 
a large-area silicon-strip detector located upstream of 
a dipole magnet with a bending power of about $4{\mathrm{\,T\,m}}$, 
and three stations of silicon-strip detectors and straw drift tubes placed downstream of the magnet.
The tracking system provides a measurement of the momentum, \ptot, of charged particles with
a relative uncertainty that varies from 0.5\% at low momentum to 1.0\% at 200\gevc.
The minimum distance of a track to a primary $pp$ collision vertex, the impact parameter (IP), is measured with a resolution of $(15+29/\pt)\mum$.
Different types of charged hadrons are distinguished using information
from two ring-imaging Cherenkov (RICH) detectors. 
Photons, electrons and hadrons are identified by a calorimeter system consisting of
scintillating-pad and preshower detectors, an electromagnetic and a hadronic calorimeter. 
Muons are identified by a system composed of alternating layers of iron and multiwire proportional chambers.
\par
The online event selection is performed by a trigger, 
which consists of a hardware stage based on information from the calorimeter and muon systems, 
followed by a software stage, which performs the charmonium candidate reconstruction.
The hardware trigger selects events with a single high transverse energy deposit in the calorimeter.
In addition, events with high multiplicity are rejected to reduce the number of random combinations of tracks (combinatorial background).
The software stage then requires two oppositely charged tracks with a good track-fit quality, identified as protons.
The studies of prompt charmonium decay to \ppbar require a dedicated software trigger selection, which suffers from a large combinatorial background. 
To suppress the combinatorial background and reduce the trigger bandwidth, the proton tracks are required to have a transverse momentum larger than 2.0\gevc, and a momentum larger than 12.5\gevc.
Charmonium candidates must have a good vertex quality and a transverse momentum larger than 5.0\gevc. 
In between the two software stages, an alignment and calibration of the detector is performed in near real-time, and their results are used in the trigger~\cite{LHCb-PROC-2015-011}. 
The same alignment and calibration information is propagated to the offline reconstruction, ensuring consistent and high-quality tracking and particle-identification information in the trigger and offline software. 
The identical performance of the online and offline reconstruction offers the opportunity to perform physics analyses directly using candidates reconstructed in the trigger \cite{LHCb-DP-2012-004,LHCb-DP-2016-001}, which the present analysis exploits. 
The storage of only the triggered candidates reduces the event size by an order of magnitude.
The signal selection is largely performed at the trigger level.
The only additional offline requirement is a tighter PID selection on the proton candidates.
To account for imperfect knowledge of the magnetic field and tracker alignment, a momentum scale calibration~\cite{LHCb-DP-2023-003} is applied to the data samples.
\par

Simulation samples are used to model the effects of the detector acceptance and the imposed selection requirements.
Generation of \proton \proton collision events is performed using the \texttt{Pythia8}~\cite{Sjostrand:2006za, Sjostrand:2007gs} event generator with specific \lhcb configuration~\cite{LHCb-PROC-2010-056}.
Decays of hadronic particles are described by EvtGen~\cite{Lange:2001uf}, in which final-state radiation is generated using PHOTOS~\cite{Golonka:2005pn}.
The interaction of generated particles with detector material is simulated using the \texttt{GEANT4} package~\cite{Agostinelli:2002hh, Allison:2006ve}. 
\par
For all simulation samples, the charmonium decay amplitude to \ppbar is modelled to be uniform in the available phase space. 
The samples of \jpsi and \psitwos mesons are simulated unpolarised. 
The samples of prompt \etac and \etactwos are generated similarly to prompt \jpsi and \psitwos, respectively, with modified mass and width according to their known values from Ref.~\cite{PDG2022}. 
A sample of \JpsiToPpbarPiz decays is generated to study the corresponding background contribution.

\section{Cross-section determination}
\label{sec:cs} 
\par
Measuring the ratios of the production cross-section of two charmonium states, a signal channel $A$ and a normalisation channel $B$, allows for a partial cancellation of several systematic effects,  
and it can be expressed as
\begin{equation}
\label{eq:csrel}
\frac{\sigma^{X}_{A}}{\sigma^{X}_{B}} = 
\frac{N^{X}_{A}}{N^{X}_{B}}\times 
\frac{\varepsilon^{X}_{B}}{\varepsilon^{X}_{A}}\times \frac{\BR_{\decay{B}{\ppbar}}}{\BR_{\decay{A}{\ppbar}}}, \\\
\end{equation}
where $X=p(\bquark)$ for prompt production (production in \bquark-hadron decays), $\sigma^{X}$ is the production cross-section, $N^{X}$ is the yield,  
and $\varepsilon^{X}$ is the total efficiency to trigger, reconstruct and select the signal and normalisation modes of the charmonium candidates. 
The branching fractions $\BR_{\decay{A(B)}{\ppbar}}$ are taken from Ref.~\cite{PDG2022}. 
The ratio of branching fractions of the inclusive \bquark-hadron decay to charmonium in Eq.~\ref{eq:csrel} is obtained as the ratio of production cross-sections in \bquark-hadron decays, $\frac{\BR_{\decay{\bquark}{AX}}}{\BR_{\decay{\bquark}{BX}}}=~\frac{\sigma^{\bquark}_{A}}{\sigma^{\bquark}_{B}}$.
The most precise determination of the \jpsi production cross-section~\cite{LHCb-PAPER-2015-037,PDG2022} is generally used to normalise the measurements. 
\par
The efficiency $\varepsilon^{X}$ is determined as the product of the geometrical acceptance, trigger, reconstruction, particle identification and selection efficiencies. 
It is calculated using simulated samples separately for prompt charmonium and charmonium from \bquark-hadron decays, and in each kinematic bin of the \etac meson for the differential production measurement.
The hardware trigger and proton PID efficiencies are corrected using data-driven methods~\cite{LHCb-DP-2018-001}.
\par
The prompt charmonium candidates (prompt sample)
and charmonium candidates from \bquark-hadron decays (\bquark-decays sample) are separated using a pseudo-decay time 
\begin{equation}
    t_z = \frac{(z_{d} - z_{p})M_{\ppbar}}{p_{z}},
 \end{equation}
where $z_{p}$ and $z_{d}$ are the positions along the beam axis of the candidate production and decay vertex, respectively, 
$M_{\ppbar}$ is the reconstructed $\ppbar$ mass and ${p_{z}}$ is the momentum component along the beam direction of a charmonium candidate. 
The prompt sample is selected by applying the requirement $t_z < 80 \fs$, while the  \bquark-decays sample is obtained by requiring $t_z > 80 \fs$ and both proton tracks to be significantly displaced from any PV.
The final result for the \etac production measurement is corrected for the cross-contamination between the prompt and \bquark-decays samples.
The correction is determined using the simulated samples, and it is calculated to be below 3\% for the prompt sample and below 2\% for the \bquark-decays sample.
This correction for the other charmonium states is negligible compared to the statistical uncertainties.

\section{Fits to data}
\label{sec:fit} 

\subsection{\texorpdfstring{Production cross-section of the \boldmath{\etactwos}, \boldmath{\hc} and \boldmath{\chicJ} mesons}{Differential production cross-section of the etac2S, hc and chicJ mesons} }
\label{subsec:fit_wide} 
\par
The yields of charmonium candidates are determined from a binned extended maximum-likelihood fit to the \ppbar mass distribution in the range 2850--3750\mevcc.
The mass fit is performed simultaneously for the prompt and \bquark-decays samples.
Independent fits are performed for studies of the \etactwos and \hc prompt production, and \chicJ production in inclusive \bquark-hadron decays.
The $J^{PC} = 1^{- -}$ charmonium states, namely the \jpsi and \psitwos mesons, have negligible natural width compared to the detector resolution, 
thus they are described by Crystal Ball functions representing the detector resolution. 
The other charmonium states are described by a convolution of a Crystal Ball function and a relativistic Breit--Wigner function. 
The \jpsi mass is a floating parameter and the \etac width is a fixed parameter in all fits. 
In the fits for the \etactwos and \hc prompt production study, masses of the \etac and \psitwos states are freely varying parameters, while masses and natural widths of the \etactwos and \hc mesons are constrained to their known values~\cite{PDG2022} in the corresponding fits. 
The masses and widths of the \chicJ states are fixed to their known values~\cite{PDG2022}.
In the \chicJ production measurement fit, masses of the \chiczero and \chicone states are allowed to vary in the fit.
Masses and widths of the other states except for the \jpsi meson are fixed to their known values~\cite{PDG2022}.
Table~\ref{tab:mass_param} summarizes the parameterisation for different fits.
\begin{table}[ht]
    \caption{Treatment of the parameters in the mass fits for the different studies. Parameters are fixed or constrained to their known values~\cite{PDG2022}.}
    \label{tab:mass_param}
    \centering
    \small
    \begin{tabular}{lcccccccc}
    \hline
         Fit        & $M_{\etac}$ & $M_{\jpsi}$ & $M_{\chiczero}$ & $M_{\chicone}$ & $M,\Gamma_{\hc}$  & $M_{\chictwo}$ & $M,\Gamma_{\etactwos}$   & $M_{\psitwos}$ \\
    \hline
         \etactwos  & Free  & Free  & Fixed     & Fixed    & Fixed       & Fixed    & Constrained & Free \\
         \hc        & Free  & Free  & Fixed     & Fixed    & Constrained & Fixed    & Fixed       & Free \\
         \chicJ     & Fixed & Free  & Free      & Free     & Fixed       & Fixed    & Fixed       & Fixed \\
    \hline
    \end{tabular}
\end{table}
One common mass resolution parameter, $\sigma_{\rm CB}$, is shared between the prompt and \bquark-decays samples.
The ratio of the mass resolution between each charmonium state and the \jpsi meson and the Crystal-Ball tail parameters are fixed according to the simulation. 
The combinatorial background is described by fourth- and third-order Chebyshev polynomial for the prompt and \bquark-decays samples, respectively.
Potential background contributions can be caused by \ppbar combinations from other partially reconstructed charmonium decays.
Only one such contribution arising from the \JpsiToPpbarPiz decay, where the \piz meson is not reconstructed, is considered in the fit. 
This background source can create a broad nonpeaking structure affecting the region below the \etac mass. 
In the fit, it is described by a square-root shape, $\sqrt{M_{\jpsi}-M_{\piz}-M_{\ppbar}}$, which is in good agreement with simulation. 
The normalisation of this background component is fixed to the \jpsi yield using known branching fractions of the \JpsiToPpbarPiz and \JpsiToPpbar decays~\cite{PDG2022}, and the efficiency ratio between the two decay modes.

\par
\begin{figure}[!tb]
  \centering
  \protect\includegraphics[width=1.00\linewidth]{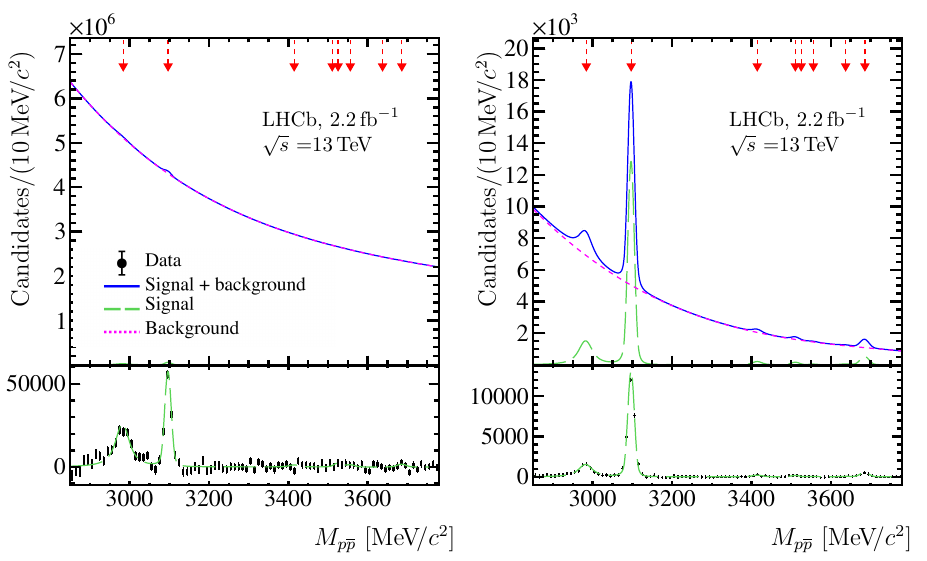}
  
  \caption{Distribution of the $p\bar{p}$ mass for (left) prompt and (right) $b$-decays samples. 
  The results of the fits using the configuration to study the $\eta_c(2S)$ prompt production are overlaid.
  Red arrows show invariant mass of charmonium states in the following order: $\eta_c$, $J/\psi$, $\chi_{c0}$, $\chi_{c1}$, $h_c(1P)$, $\chi_{c2}$, $\eta_c(2S)$ and $\psi(2S)$ states.
  The lower panel shows the data and the fit model with the background subtracted.} 
  \label{fig:massfit}
\end{figure}

The \ppbar mass distribution for the prompt and \bquark-decays samples is shown in Fig.~\ref{fig:massfit}.
Projections of the simultaneous fit result are overlaid. 
The yields of the charmonium states from the fit are listed in Table~\ref{tab:yields_sim}.
The fit shows no significant \etactwos and \hc signals. 
Therefore, for these states, upper limits are set on the relative \etactwos to \jpsi and \hc to \jpsi production cross-section at 90\% and 95\% confidence levels (CL). 

\begin{table}[!t] 
  \caption{Results of the simultaneous fit to the \ppbar mass distributions for combined prompt and \bquark-decays data sample. Yields are restricted to be positive with symmetric uncertainties presented. The uncertainties are statistical only. }
  \centering
  {\small{
  \begin{tabular}{lr@{\:$\pm$\:}lr@{\:$\pm$\:}lr@{\:$\pm$\:}l} 
  \hline
  Charmonium state        & \multicolumn{2}{c}{$N^{p}\times 10^{2}$}            & \multicolumn{2}{c}{$N^{b}$}  & \multicolumn{2}{c}{Mass [\mevcc\!]}  \\ \hline
  \etac       & 1501 & 99   & 8857  & 361  & 2982.1  &  0.8 \\
  \jpsi       & 1225 & 42   & 27983 & 248  & 3096.6  & 0.1\\
  \chiczero   & 47   & 50   & 766   & 136  &  \multicolumn{2}{c}{3414.7} \\
  \chicone    & $0 $ & 62   & 443   & 98   &  \multicolumn{2}{c}{3510.7}  \\
  \hc         & 63   & 37   & $0 $  & 67   &  \multicolumn{2}{c}{3525.4}\\
  \chictwo    & 38   & 39   & 125   & 97   &  \multicolumn{2}{c}{3556.2}\\
  \etactwos   & $0 $ & 32   & $0 $  & 52   &  \multicolumn{2}{c}{3637.5} \\
  \psitwos    & 49   & 37   & 1529  & 91   & 3685.3  &  0.8 \\
  \hline
  \end{tabular}
  }}
  \label{tab:yields_sim}
\end{table}
\par

\subsection{\texorpdfstring{Differential production cross-section of the \boldmath{\etac} meson}{Differential production cross-section of the etac meson} }
\label{subsec:fit_etac} 
\par
The \etac differential production cross-section measurement is performed in bins of the \etac transverse momentum and rapidity.
To determine the ratio of the \etac to \jpsi yields, the fit to the \ppbar mass distribution is performed simultaneously in bins of either \pt or \y.
The range of the mass fit is reduced to 2850--3250\mevcc.
The model of the signal description is the same as that for the fits described above.
The dependency of the mass resolution on \pt and \y is determined from simulation and scaled by a factor in the fit to data.
The scaling parameter is obtained from a fit to the mass distribution in the simulated samples using the same signal parameterisation as that for data.
The \jpsi mass and the mass difference between the \jpsi and \etac states are left as floating parameters in the fit.
The combinatorial background is described by an exponential function multiplied by a second-order polynomial for both prompt and \bquark-decays samples.
The cross-contamination between the samples is estimated from simulation in bins of \pt and \y, and used for correcting the yields of the charmonium states.
\par
The relative \etac to \jpsi yield is extracted from the fit and listed in Table~\ref{tab:yields_pt} and Table~\ref{tab:yields_y} for \pt and \y bins, respectively.
\begin{table}[!b] 
  \caption{\small{Ratios of the observed \etac to \jpsi yields corrected for cross-contamination for prompt production and production in \bquark-hadron decays in bins of transverse momentum \pt. The uncertainties are statistical.}}
  \centering
  {\small{
  \begin{tabular}{r@{\:$-$\:}lcc} 
  \hline
  \multicolumn{2}{l}{\pt [\gevc]}
       & $N^{p}_{\etac}/N^{p}_{\jpsi}$            & $N^{\bquark}_{\etac}/N^{\bquark}_{\jpsi}$    \\ \hline
  5.0 & 6.5      & $0.34\pm0.32$ 	& $0.295\pm0.057$ \\
  6.5 & 8.0      & $0.70\pm0.21$ 	& $0.247\pm0.034$ \\
  8.0 & 10.0     & $1.07\pm0.15$ 	& $0.302\pm0.027$ \\
  10.0 & 12.0    & $0.88\pm0.20$ 	& $0.358\pm0.034$ \\
  12.0 & 14.0    & $1.34\pm0.29$ 	& $0.318\pm0.039$ \\
  14.0 & 20.0    & $1.30\pm0.42$ 	& $0.313\pm0.042$ \\
  \hline
  5.0 & 20.0     & $0.87\pm0.09$ 	& $0.303\pm0.017$ \\
  \hline
  \end{tabular}
  }}
  \label{tab:yields_pt}
\end{table} 

\begin{table}[ht] 
  \caption{\small{Ratios of the observed \etac to \jpsi yields mesons corrected for cross-contamination for prompt production and production in \bquark-hadron decays in bins of rapidity \y. The uncertainties are statistical.}}
  \centering
  {\small{
  \begin{tabular}{lcc} 
  \hline
  \y       & $N^{p}_{\etac}/N^{p}_{\jpsi}$            & $N^{\bquark}_{\etac}/N^{\bquark}_{\jpsi}$    \\ \hline
  2.0 -- 2.5     & $0.64\pm0.15$ 	& $0.304\pm0.027$ \\
  2.5 -- 3.0     & $1.12\pm0.19$ 	& $0.315\pm0.027$ \\
  3.0 -- 3.5     & $0.86\pm0.19$ 	& $0.300\pm0.029$ \\
  3.5 -- 4.0     & $0.83\pm0.40$ 	& $0.317\pm0.066$ \\
  \hline
  \end{tabular}
  }}
  \label{tab:yields_y}
\end{table} 

\section{Systematic uncertainties}
\label{sec:syst} 
\par
Alternative fit parameterisations for the signal and background are used to estimate systematic uncertainties on the production cross-section. 
For the study of the \etactwos and \hc prompt production, the systematic uncertainties are included in the upper limit using the discrete profiling method~\cite{Dauncey:2014xga}. 
For the other measurements, the systematic uncertainties are estimated as a difference between the baseline fit and the fit with the alternative parameterisation.
The following systematic uncertainties are common for all measurements.

\par 
The uncertainty associated with the mass resolution description is estimated using an alternative fit where the sum of two Gaussian functions with parameters determined from simulation is used instead of the Crystal Ball function. 
The uncertainty related to the knowledge of the charmonium natural width is estimated in alternative fits where the natural width is varied within the world-average uncertainties~\cite{PDG2022}. 
The largest variation is taken as the systematic uncertainty.
The uncertainty associated with the combinatorial background description is estimated using an alternative background parameterisation. 
In the studies of the \etactwos and \hc prompt production cross-section and \chicJ production in inclusive \bquark-hadron decays, the background is parametrised using an exponential function multiplied by a third- or a second-order polynomial function for prompt and \bquark-decays samples, respectively.
In addition, in the \etactwos and \hc prompt production study, a second parameterisation uses the product of a
square-root function, an exponential function and a polynomial. 
A third-order polynomial is used for the prompt sample, and a second-order polynomial function is used for the \bquark-decays sample.
In the \etac production studies, an alternative background parameterisation with a third-order Chebyshev polynomial is used.
The systematic uncertainty related to the contribution from \JpsiToPpbarPiz decays is estimated by varying the value of the efficiency ratio $\varepsilon_{\JpsiToPpbarPiz}/\varepsilon_{\JpsiToPpbar}$ and the branching ratio $\BR_{\JpsiToPpbarPiz}/\BR_{\JpsiToPpbar}$ within their uncertainties.

\par
The systematic uncertainty corresponding to the ratio of efficiencies of different charmonium states is estimated by varying its value by the uncertainty corresponding to the simulation sample sizes.
The uncertainties associated with possible discrepancies between data and simulation are largely cancelled out in the efficiency ratio.
The uncertainty related to the cross-contamination between the prompt and \bquark-decays samples is estimated by modifying the corresponding yields by their uncertainties. 

\par
The effect of possible nonzero polarisation of the \jpsi meson is accounted for in the reconstruction efficiency. 
Weights are assigned to the simulated \jpsi \to \ppbar events to reproduce the distributions corresponding to the polarisation parameter values of $\lambda_{\theta}=\pm 0.1$~\cite{LHCb-PAPER-2015-037}.

\par
The uncertainty of the efficiency ratios, as well as the uncertainties of the $\BR_{\decay{\jpsi}{\ppbar}}$ and $\BR_{\decay{\etac}{\ppbar}}$ values, are taken into account directly in the production cross-section determination.
The systematic uncertainties on the signal yields are propagated to the production cross-section using Eq.~\ref{eq:csrel}.

\par
The \etac differential production cross-section measurement includes the uncertainty corresponding to the \pt or \y dependence of the relative mass resolution between \etac and \jpsi states. 
It is accounted for by introducing a linear dependence of the resolution as a function of \pt or \y constrained by simulation. 
For the \etac differential production cross-section measurement, the systematic uncertainties on the combinatorial background description, resolution ratio, cross-contamination between samples and the contribution from the \JpsiToPpbarPiz decay are estimated bin-by-bin and considered as uncorrelated among bins.
To reduce the impact of statistical fluctuations, they are smoothed over bins using a linear interpolation.
The systematic uncertainties on the \etac natural width, mass-resolution model and \jpsi polarisation are considered as correlated among bins.
The first two are estimated from the fit to the mass distribution in the entire \pt(\y) range and assumed to be the same in all the bins.
The uncertainties on the relative \etac prompt production cross-sections are given in Table~\ref{tab:uncert_tot}.
The uncertainties in bins of \pt and \y are summarised in Tables~\ref{tab:uncert_pt} and~\ref{tab:uncert_y} in the Appendix.

\begin{table}[!t]
    \centering \small
   \caption{\small{Relative uncertainties (in \%) on the ratio of prompt cross-sections $\sigma^{p}_{\etac}/\sigma^{p}_{\jpsi}$.
    Uncertainties on $\BR_{\etac\to\ppbar}$ and $\BR_{\jpsi\to\ppbar}$ measurements are considered separately and given in the text.}}
    \label{tab:uncert_tot}
    \begin{tabular}{lc} \hline
\multirow{ 2}{*}{}  & \phantom{00}\pt [\gevc] \\ \cmidrule{2-2} 
                    & \phantom{00}5.0 -- 20.0 \\ \hline
Statistical uncertainty                     
& \phantom{00}10.4 	\\ \hline
Combinatorial background               
& \phantom{00}5.4 	\\ 
Contribution from \JpsiToPpbarPiz     
& \phantom{00}0.3 	\\ 
Resolution \pt-dependence  
& \phantom{00}0.1 	\\ 
Cross-feed                            
& $<0.1$ 	\\ 
Total uncorrelated systematic
& \phantom{00}5.4 	\\ \hline
Variation of $\Gamma_{\etac}$         
& \phantom{00}1.7 	\\ 
Mass resolution model                 
& \phantom{00}1.8 	\\ 
Polarisation of \jpsi                 
& \phantom{00}1.6 	\\ 
Total correlated systematic                     
& \phantom{00}3.0 	\\ \hline
Total systematic                            
& \phantom{00}6.2 	\\ \hline 
    \end{tabular}
 \end{table}
\section{Results}
\label{sec:results}

\subsection{\texorpdfstring{Prompt production of the \boldmath{\etactwos} and \boldmath{\hc} mesons}{etac2S and hc production}}
\label{subsec:etac2SCS}

\par
\begin{figure}[ht]
    \centering
    \protect\includegraphics[width=1.00\linewidth]{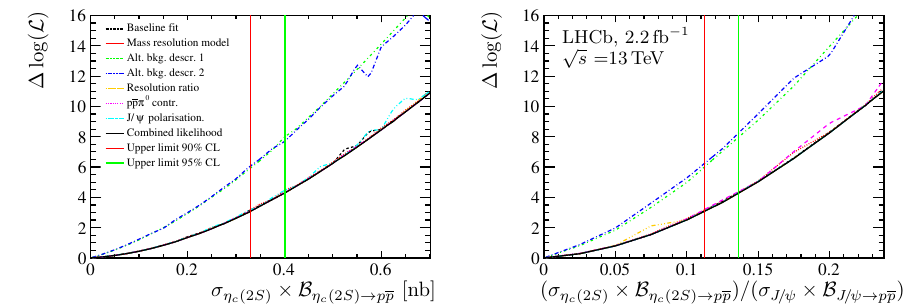}
    
    \caption{\small{
    The log-likelihood profile plots for the (black dashed line) baseline fit and alternative fits including: (red dashed line) alternative mass resolution model, 
    (green and blue dashed lines) alternative models of combinatorial background, (orange dashed line) alternative resolution ratio, 
    (magenta dashed line) alternative contribution from partially reconstructed $J/\psi\to p\bar{p}\pi^{0}$ decay, and (cyan dashed line) accounting for the nonzero $J/\psi$ polarisation.
    The black solid line represents the log-likelihood profile obtained using the discrete profiling method~\cite{Dauncey:2014xga}.
    The solid red and green lines show Bayesian upper limits on (left) $\eta_c(2S)$ and (right) relative $\eta_c(2S)$ to $J/\psi$ prompt-production cross-section at 90\% and 95\% CL, respectively.}}
    \label{fig:ULBay_etac2S}
\end{figure}
\begin{figure}[ht]
    \centering
    \protect\includegraphics[width=1.00\linewidth]{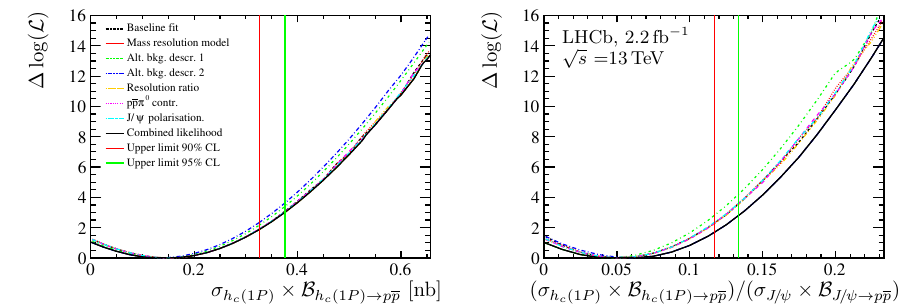} 
       
    \caption{\small{
    The log-likelihood profile plots for the (black dashed line) baseline fit and alternative fits including: 
    (red dashed line) alternative mass resolution model, (green and blue dashed lines) alternative models of combinatorial background, 
    (orange dashed line) alternative resolution ratio, (magenta dashed line) alternative contribution from partially reconstructed $J/\psi\to p\bar{p}\pi^{0}$ decay, and (cyan dashed line) accounting for the nonzero $J/\psi$ polarisation.
    The black solid line represents the log-likelihood profile obtained using the discrete profiling method~\cite{Dauncey:2014xga}.
    The solid red and green lines show Bayesian upper limits on (left) $h_c(1P)$ and (right) relative $h_c(1P)$ to $J/\psi$ prompt-production cross-section at 90\% and 95\% CL, respectively.
    }} 
    \label{fig:ULBay_hc}
\end{figure}

The upper limits at 90\% and 95\% confidence level (CL) on the relative and absolute prompt production cross-section at $\sqs = 13\tev$ for the \etactwos and \hc states are set for the first time.
The study is performed in the kinematic range $5.0<\pt<20.0\gevc$ and  $2.0<y<4.0$.
The results are listed in Table~\ref{tab:UL_comp} and shown in Figs.~\ref{fig:ULBay_etac2S} and ~\ref{fig:ULBay_hc}. 
The Bayesian approach~\cite{DAgostini:2003bpu} was used for setting the upper limits.
\begin{table}[ht] 
    \caption{\small{Upper limits at 90\% and 95\% CL for the \etactwos and \hc prompt production cross-sections and their ratios to that of the \jpsi meson, extracted using a Bayesian approach.}}
    \centering
    {\small{
    \begin{tabular}{lcc} 
    \hline
     Upper limit & $90\%$CL  & $95\%$CL  \\ \hline
     $(\sigma_{\etactwos} \times \BR_{\etactwos\to\ppbar})/ (\sigma_{\jpsi} \times \BR_{\jpsi\to\ppbar})$
     & 0.113 & 0.136 \\
     $\sigma_{\etactwos} \times \BR_{\etactwos\to\ppbar}\ [\rm{nb}]$
     & 0.331 & 0.401  \\
     $(\sigma_{\hc} \times \BR_{\hc\to\ppbar})/ (\sigma_{\jpsi} \times \BR_{\jpsi\to\ppbar})$
     & 0.117 & 0.133  \\
     $\sigma_{\hc} \times \BR_{\hc\to\ppbar}\ [\rm{nb}]$
     & 0.327 & 0.375 \\
     \hline
    \end{tabular}
    }}
    \label{tab:UL_comp}
\end{table}

\par
A comparison of the results for the \etactwos production and an NLO NRQCD theoretical prediction~\cite{Lansberg:2017ozx} is shown in Table~\ref{tab:etac2S_ul_theory}.
The comparison is performed in different kinematic ranges due to the lack of \jpsi prompt production cross-section measurement in the same kinematic range as the current \etactwos measurement.
Therefore, an additional upper limit for the \etactwos prompt production cross-section at 95\% CL is set in the \pt range $5.0<\pt<14.0\gevc$.
Assuming that the difference between the \jpsi prompt production cross-section values estimated in the kinematic range $5.0<\pt<14.0\gevc$ and $5.0<\pt<20.0\gevc$ is compatible with their uncertainty, the absolute \etactwos production cross-section in both \pt ranges is estimated using the first value.
The predictions are made using three different sets of LDMEs:  Shao et al. \cite{Shao:2014yta}, Gong et al. \cite{Gong:2012ug}, and Bodwin et al. \cite{Bodwin:2015iua}.
The upper limit is found to be considerably lower than the prediction based on Bodwin et al.\cite{Bodwin:2015iua}. 
The obtained cross-section upper limit is also below the prediction by Shao et al. \cite{Shao:2014yta}, which, however, reports large theory uncertainties.

\begin{table}[!t] 
\caption{\small{Upper limits (UL) at 95\% CL for \etactwos relative to \jpsi and \etactwos absolute prompt production cross-section, and its comparison with NLO NRQCD theoretical prediction (Ref.~\cite{Lansberg:2017ozx,Shao:2014yta,Gong:2012ug,Bodwin:2015iua}).}}
\centering
{\small{
    \begin{tabular}{lcccc} 
     \hline
    \pt [\gevc] & UL $95\%\text{CL}$ & Shao et al. \cite{Shao:2014yta} & Gong et al. \cite{Gong:2012ug} & Bodwin et al. \cite{Bodwin:2015iua} \\ \hline
    & \multicolumn{4}{l}{$\sigma_{\etactwos} \times \BR_{\etactwos\to\ppbar}\ [\rm{pb}]  $}  \\\cmidrule{2-5}
    5.0 -- 14.0 & $<426$ &  $ 664 \pm 297$ & $ 365\pm 135$ & $ 855 \pm 123$  \\
    5.0 -- 20.0 & $<401$ &  $ 674 \pm 304$ & $ 368\pm 138$ & $ 870 \pm 126$  \\ \hline
    & \multicolumn{4}{l}{$(\sigma_{\etactwos} \times \BR_{\etactwos\to\ppbar})/ (\sigma_{\jpsi} \times \BR_{\jpsi\to\ppbar}) $}  \\ \cmidrule{2-5}
    5.0 -- 14.0, \y$<$4.0 & $<0.14$ &  $0.48\pm0.22$ & $0.27\pm0.10$ & $0.62\pm0.09$  \\
    5.0 -- 14.0, \y$<$4.5 &         &  $0.43\pm0.19$ & $0.24\pm0.09$ & $0.55\pm0.08$  \\
    5.0 -- 20.0, \y$<$4.0 & $<0.14$ & & & \\
    \hline
\end{tabular}
}}
\label{tab:etac2S_ul_theory}
\end{table}

\subsection{\texorpdfstring{Production of the \boldmath{\etac} meson}{etac production}}
\label{subsec:etacCS}

\par
The differential production cross-section of the \etac meson is measured for both prompt charmonium production and production in inclusive \bquark-hadron decays.
Results are presented in an extended \pt range in comparison to the previous analysis \cite{LHCb-PAPER-2019-024}.
The \pt-differential cross-section is measured in the range $5.0<\pt<20.0\gevc$, and \y-differential cross-section is measured for the first time in the range $2.0<\y<4.0$.

\par 
Using Eq.~\ref{eq:csrel} and yields extracted from the fit to data, the relative \etac to \jpsi prompt production cross-section is measured to be
\begin{equation*}
  (\sigma_{\etac}/\sigma_{\jpsi})^{5.0<\pt<20.0\gevc, \ 2.0<\y<4.0} = 1.32 \pm 0.14 \pm 0.09 \pm 0.13.  
\end{equation*}
Here and throughout the paper, unless explicitly mentioned, the first uncertainty is statistical, the second is systematic, and the third is due to the uncertainties on the branching fractions $\BR_{\EtacToPpbar}$ and $\BR_{\JpsiToPpbar}$.

The prompt \jpsi production cross-section is only available for $\pt<14\gevc$.
Using the value of $\sigma_{\jpsi}^{5.0<\pt<14.0\gevc, \ 2.0<\y<4.0}=1373 \pm 58 \nb$ from Ref.~\cite{LHCb-PAPER-2015-037}, the absolute \etac prompt production cross-section is measured to be 
\begin{equation*}
  (\sigma_{\etac})^{5.0<\pt<14.0\gevc, \ 2.0<\y<4.0} = 1815 \pm 189 \pm 120 \pm 192 \nb,  
\end{equation*}
where the last uncertainty also includes the uncertainty on the \jpsi production cross-section.

\begin{figure}[!t]
    \centering
    \protect\includegraphics[width=1.00\linewidth]{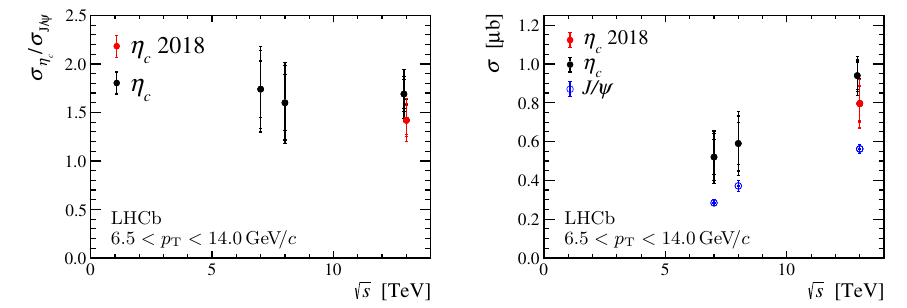}
    
    \caption{\small{Left: measured $\eta_c$ to $J/\psi$ cross-section ratio at different centre-of-mass energies. 
    Right: measured  $\eta_c$ and $J/\psi$~\cite{LHCb-PAPER-2013-008,LHCb-PAPER-2013-016,LHCb-PAPER-2015-037} prompt production cross-sections.
    The red points show the result from the current analysis.
    The error bars show following uncertainties: statistical, systematic, and due to the $J/\psi\to p\bar{p}$ and $\etac\to p\bar{p}$ branching fractions (and $J/\psi$ production cross-section for the absolute production cross-section).
    } }
    \label{fig:CSSqs}
\end{figure}

The relative \etac to \jpsi and absolute \etac prompt production cross-section are shown as a function of the centre-of-mass energy in Fig.~\ref{fig:CSSqs}.
In addition, the corresponding \jpsi production cross-section in the same kinematic range from Refs.~\cite{LHCb-PAPER-2013-008,LHCb-PAPER-2013-016,LHCb-PAPER-2015-037} is shown for comparison.

The relative \etac inclusive branching fraction from \bquark-hadron decays is measured to be
\begin{equation*}
  \BR_{\bquark\to\etac X}/\BR_{\bquark\to\jpsi X} = 0.49 \pm 0.03 \pm 0.02 \pm 0.05.
\end{equation*}
Using the value $\BR_{\bquark\to\jpsi X} = 1.16\pm 0.10\%$~\cite{PDG2022}, the absolute value is measured to be
\begin{equation*}
    \BR_{\bquark\to\etac X} = (5.64 \pm 0.31 \pm 0.18 \pm 0.73)\times 10^{-3},
\end{equation*} 
where the last uncertainty includes the uncertainty on the $\bquark\to\jpsi$ inclusive branching fraction.

\par
\begin{figure}[ht!]
    \centering
    \protect\includegraphics[width=1.00\linewidth]{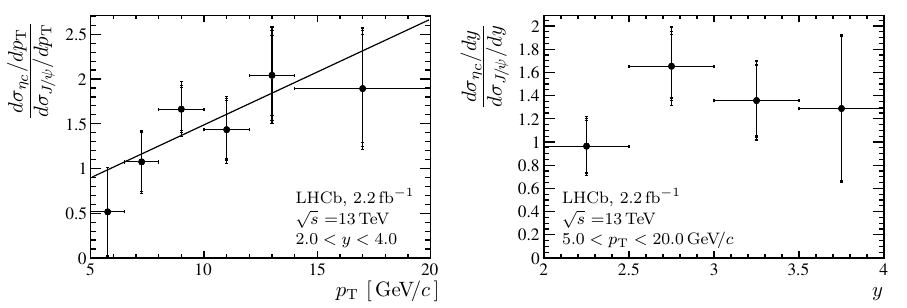}
    
    \caption{\small{Measured $\eta_c$ to $J/\psi$ prompt-production cross-section ratio in intervals of (left) $p_{\rm{T}}$ and (right) $y$.
    The error bars show following uncertainties: statistical, systematic, and due to the branching fractions $\BR_{J/\psi\to p\bar{p}}$ and $\BR_{\etac\to p\bar{p}}$ uncertainties. 
    The result of the fit with a linear function for $p_{\rm{T}}$-differential cross-section is overlaid.}
    } 
    \label{fig:CSRel}
\end{figure}

The relative differential \etac to \jpsi prompt production cross-section is shown in Fig.~\ref{fig:CSRel} in bins of \pt and \y, and 
the corresponding values are reported in Tables~\ref{tab:relCSPT} and~\ref{tab:relCSY}, respectively.
The result of the fit by a linear function to the prompt production cross-section in bins of \pt is overlaid. 
The slope is found to be $0.118 \pm  0.057 (\gevc^{-1})$. 
A significant slope may indicate the presence of CO contribution to the \etac prompt production cross-section at higher \pt~\cite{Han:2014jya}.

\par
\begin{figure}[!t]
    \centering
    \protect\includegraphics[width=1.00\linewidth]{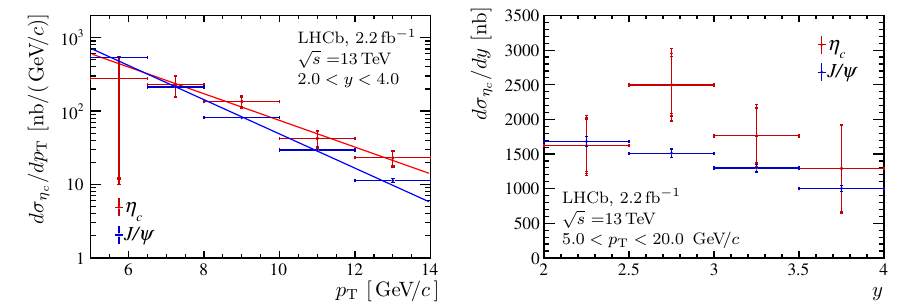}
    
    \caption{\small{Measurement of the differential prompt production cross-section of the (red) $\eta_c$ and (blue) $J/\psi$ mesons~\cite{LHCb-PAPER-2015-037} in bins of (left) $p_{\rm{T}}$- and (right) $y$. 
    The error bars show following uncertainties for $\eta_c$ production: statistical, systematic, and the uncertainty due to the $J/\psi\to p\bar{p}$ and $\etac\to p\bar{p}$ branching fractions and $J/\psi$ production cross-section. 
    The results of the fits with exponential functions are overlaid.}
    } 
    \label{fig:CSAbs}
\end{figure}

The absolute differential \etac prompt production cross-sections as a function of \pt and \y are shown in Fig.~\ref{fig:CSAbs}. 
The values are reported in Tables~\ref{tab:absCSPT} and~\ref{tab:absCSY} in bins of \pt and \y, respectively.

\par
\begin{figure}[!t]
    \centering
    \protect\includegraphics[width=1.00\linewidth]{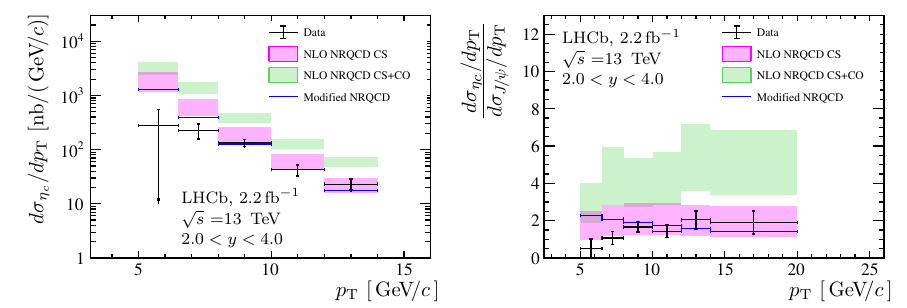}
    \caption{\small{Measurement of the (left) absolute $\eta_c$ and (right) relative $\eta_c$ to $J/\psi$ $p_{\rm{T}}$-differential prompt-production cross-section. 
    Magenta boxes represent NLO NRQCD CS prediction; green boxes show the sum of NLO NRQCD CS and CO predictions.
    Blue lines represent the modified NRQCD prediction~\cite{Biswal:2023mwu}.}
    } 
    \label{fig:CS_theory}
\end{figure}

Relative and absolute prompt \pt-differential production results are compared with the NLO NRQCD predictions for CS and for the sum of CS and CO contributions~\footnote{The prediction is provided by H.S.Shao  based on Ref.~\cite{Han:2014jya} in the kinematic range corresponding to that presented in this analysis.} and modified NRQCD~\cite{Biswal:2023mwu}, and shown in Fig.~\ref{fig:CS_theory}. 
The NRQCD CS model and modified NRQCD show reasonable agreement with data for $\pt>8\gevc$, but overestimate data at lower \pt. 
The the sum of the CS and CO contributions overestimates data in the entire \pt range. 

\subsection{\texorpdfstring{Production of \boldmath{\chicJ} mesons in \bquark-hadron decays}{The chicJ production in \bquark-hadron decays}}
\label{subsec:chicJBR} 
The relative yields of the \chicJ states in \bquark-hadron inclusive decays are measured in order to derive the absolute branching ratios of the \chicJ states to~\ppbar.

\par
The relative efficiency-corrected yields of the \chicJ states in \bquark-hadron inclusive decays are measured to be
\begin{align*}
    \frac{\BR(\decay{\bquark}{\chicone X})\times\BR(\decay{\chicone}{\ppbar})}{\BR(\decay{\bquark}{\chiczero X})\times\BR(\decay{\chiczero}{\ppbar})}  &= 0.58 \pm 0.23 \pm 0.02 \, , \\
    \frac{\BR(\decay{\bquark}{\chictwo X})\times\BR(\decay{\chictwo}{\ppbar})}{\BR(\decay{\bquark}{\chiczero X})\times\BR(\decay{\chiczero}{\ppbar})}  &= 0.17 \pm 0.12 \pm 0.01 \, ,
\end{align*}
where the quoted uncertainties are statistical and systematic. The uncertainty in both measurements is dominated by the statistical uncertainty.
\par
Using the known branching ratios of the \chicJ decays to the \ppbar final state, ${\BR_{\decay{\chiczero}{\ppbar}} = (2.21 \pm 0.8)\times 10^{-4}}$, ${\BR_{\decay{\chicone}{\ppbar}} = (7.60 \pm 0.34)\times 10^{-5}}$, and ${\BR_{\decay{\chictwo}{\ppbar}} = (7.33 \pm 0.33)\times 10^{-5}}$~\cite{PDG2022}, the ratios of the branching fractions of the \bquark-hadron inclusive decays to the \chicJ states are measured to be
\begin{align*}
    \frac{\BR(\decay{\bquark}{\chicone X})}{\BR(\decay{\bquark}{\chiczero X})}  &= 1.68 \pm 0.66 \pm 0.05 \pm 0.10 \, , \\
    \frac{\BR(\decay{\bquark}{\chictwo X})}{\BR(\decay{\bquark}{\chiczero X})}  &= 0.50 \pm 0.36 \pm 0.04 \pm 0.03 \, ,
\end{align*}
where the quoted uncertainties are statistical, systematic and from the branching fractions of the \chicJ decays to \ppbar final state, respectively.
\par
Precise measurements of the branching ratios of the \jpsi decay to the \ppbar final state and the inclusive \bquark-hadron decay to the \jpsi final state enable an improved precision of the measurements for the \chiczero and \chicone states.
The product of the branching fractions of \bquark-hadron decays to \chicJ and the branching fractions of the \chicJ\to\ppbar decay mode are determined as
\begin{align*}
    \BR_{\decay{\bquark}{\chiczero X}}\times\BR_{\decay{\chiczero}{\ppbar}} &= (6.74 \pm 1.18 \pm 0.18 \pm 0.59)\times 10^{-7} \, , \\
    \BR_{\decay{\bquark}{\chicone X}} \times\BR_{\decay{\chicone}{\ppbar}}  &= (3.88 \pm 0.91 \pm 0.11 \pm 0.34)\times 10^{-7} \, , \\
    \BR_{\decay{\bquark}{\chictwo X}} \times\BR_{\decay{\chictwo}{\ppbar}}  &= (1.13 \pm 0.83 \pm 0.03 \pm 0.10)\times 10^{-7} \, .
\end{align*}
The quoted uncertainties are statistical, systematic and related to the $\BR_{\decay{\jpsi}{\ppbar}}$, and $\BR_{\decay{\bquark}{\jpsi X}}$ uncertainties.
The computed branching ratios of inclusive \bquark-hadron decays to \chicJ are presented in Table~\ref{tab:chicJBR}.
\par
The values are compared with the previous \lhcb measurement of \chicJ production in \bquark-hadron decays~\cite{LHCb-PAPER-2017-007} and with the world averages~\cite{PDG2022}.
Both \lhcb results are in good agreement.
The presented measurements of the \chiczero and \chicone production in \bquark-hadron decays are the most precise to date. 
\begin{table}[!t]
    \centering
    \small
    \caption{\small{Branching fractions of \chicJ production in inclusive \bquark-hadron decays. The quoted uncertainties are statistical, systematic and the related to the $\BR_{\decay{\chicJ}{\ppbar}}$, $\BR_{\decay{\jpsi}{\ppbar}}$, and $\BR_{\decay{\bquark}{\jpsi X}}$ uncertainties. The values are compared with results from analysis of charmonia production using decays to $\phi\phi$~\cite{LHCb-PAPER-2017-007} and the world average values~\cite{PDG2022}.}}
    \label{tab:chicJBR}
    \begin{tabular}{lccr@{\:$\pm$\:}l}
    \hline
        & $\ccbar\to\ppbar$, measured & $\ccbar\to\phi\phi$~\cite{LHCb-PAPER-2017-007} & \multicolumn{2}{c}{World average\cite{PDG2022}} \\
    \hline
     $\BR_{\decay{\bquark}{\chiczero X}}\times10^{-3}$ 
        & $3.05 \pm 0.54 \pm 0.08 \pm 0.29$ & $3.02\pm1.08$ & 15 & 6\\
     $\BR_{\decay{\bquark}{\chicone X}}\times10^{-3}$ 
        & $5.11 \pm 1.20 \pm 0.14 \pm 0.50$ & $2.76\pm1.09$ & 14 & 4\\
     $\BR_{\decay{\bquark}{\chictwo X}}\times10^{-3}$ 
        & $1.54 \pm 1.13 \pm 0.04 \pm 0.15$ & $1.15\pm0.42$ & 6.2 & 2.9\\
     \hline
    \end{tabular}

\end{table}
\par    

\section{Summary}
\label{sec:summary}

Using \proton\proton collision data corresponding to an integrated luminosity of $2.2 \invfb$ collected by the \lhcb experiment at $\sqs=13\tev$ in 2018, 
the upper limits on the ratio of the \etactwos and \hc to \jpsi  prompt production cross-section, at 90\% (95\%) CL, are determined for the first time as
\begin{equation*}
    \frac{(\sigma_{\etactwos} \times \BR_{\etactwos\to\ppbar})} {(\sigma_{\jpsi} \times \BR_{\jpsi\to\ppbar})} < 0.11\ (0.14),
\end{equation*}
\begin{equation*}
    \frac{(\sigma_{\hc} \times \BR_{\hc\to\ppbar})} {(\sigma_{\jpsi} \times \BR_{\jpsi\to\ppbar})} < 0.12\ (0.13),
\end{equation*}
in the kinematic range $5.0<\pt<20.0\gevc$ and $2.0<\y<4.0$.
The upper limits on the \etactwos and \hc prompt production cross-sections multiplied by their decay branching fractions to the \ppbar final state are determined. 

The relative \etac to \jpsi prompt production cross-section is measured in the extended kinematic range compared to the previous analysis, $5.0<\pt<20.0\gevc$, leading to
\begin{equation*}
    (\sigma_{\etac}/\sigma_{\jpsi})^{5.0< \pt < 20.0\gevc,\ 2.0<\y<4.0} = 1.32 \pm 0.14 \pm 0.09 \pm 0.13.
\end{equation*} 
Henceforth, the first uncertainty is statistical, and the second is systematic. 
The third uncertainty is due to the uncertainties on the branching fractions $\BR_{\EtacToPpbar}$ and $\BR_{\JpsiToPpbar}$.

The absolute \etac prompt production cross-section is derived using the \jpsi prompt production cross-section measurement at \sqs=13\tev~\cite{LHCb-PAPER-2015-037}.
The relative \etac to \jpsi and the absolute \etac inclusive branching fractions from \bquark-hadron decays are measured, obtaining
\begin{equation*}
    \BR_{\bquark\to\etac X}/\BR_{\bquark\to\jpsi X} = 0.49 \pm 0.03 \pm 0.02 \pm 0.05,
\end{equation*} 
\begin{equation*}
    \BR_{\bquark\to\etac X} = (5.64 \pm 0.31 \pm 0.18 \pm 0.73)\times 10^{-3},
\end{equation*} 
where the third uncertainty is due to the uncertainties on the branching fractions $\BR_{\EtacToPpbar}$ and $\BR_{\JpsiToPpbar}$ and the uncertainties on the branching fractions $\BR_{\EtacToPpbar}$, $\BR_{\JpsiToPpbar}$ and $\BR_{\bquark\to\jpsi X}$, respectively.
The results on \etac production are consistent with the previous \lhcb measurement at \sqs=13\tev \cite{LHCb-PAPER-2019-024}. 
The \etac prompt production cross-section at low \pt is below the available theoretical expectations and calls for an improved theory description in this range. 
\par
New measurements of the branching fractions $\BR_{\decay{\bquark}{\chicJ X}}\times\BR_{\decay{\chicJ}{\ppbar}}$ and $\BR_{\decay{\bquark}{\chicJ X}}$ 
are reported.
The inclusive branching fraction results for the \chiczero and \chicone mesons, 
\begin{align*}
    \BR_{\decay{\bquark}{\chiczero X}} &= (3.05 \pm 0.54 \pm 0.08 \pm 0.29) \times10^{-3}, \\
    \BR_{\decay{\bquark}{\chicone X}} &= (5.11 \pm 1.20 \pm 0.14 \pm 0.50) \times10^{-3}, \\
\end{align*}
are the most precise to date. The third uncertainty here is due to the uncertainties on the branching fractions $\BR_{\chi_{c0,1}\to\ppbar}$, $\BR_{\JpsiToPpbar}$ and $\BR_{\bquark\to\jpsi X}$.





\section*{Acknowledgements}
%
%
\noindent We thank J.P. Lansberg, H.S. Shao and K. Sridhar for useful discussions and for providing predictions for charmonium production.
We express our gratitude to our colleagues in the CERN
accelerator departments for the excellent performance of the LHC. We
thank the technical and administrative staff at the LHCb
institutes.
We acknowledge support from CERN and from the national agencies:
CAPES, CNPq, FAPERJ and FINEP (Brazil); 
MOST and NSFC (China); 
CNRS/IN2P3 (France); 
BMBF, DFG and MPG (Germany); 
INFN (Italy); 
NWO (Netherlands); 
MNiSW and NCN (Poland); 
MCID/IFA (Romania); 
MICIU and AEI (Spain);
SNSF and SER (Switzerland); 
NASU (Ukraine); 
STFC (United Kingdom); 
DOE NP and NSF (USA).
We acknowledge the computing resources that are provided by CERN, IN2P3
(France), KIT and DESY (Germany), INFN (Italy), SURF (Netherlands),
PIC (Spain), GridPP (United Kingdom), 
CSCS (Switzerland), IFIN-HH (Romania), CBPF (Brazil),
and Polish WLCG (Poland).
We are indebted to the communities behind the multiple open-source
software packages on which we depend.
Individual groups or members have received support from
ARC and ARDC (Australia);
Key Research Program of Frontier Sciences of CAS, CAS PIFI, CAS CCEPP, 
Fundamental Research Funds for the Central Universities, 
and Sci. \& Tech. Program of Guangzhou (China);
Minciencias (Colombia);
EPLANET, Marie Sk\l{}odowska-Curie Actions, ERC and NextGenerationEU (European Union);
A*MIDEX, ANR, IPhU and Labex P2IO, and R\'{e}gion Auvergne-Rh\^{o}ne-Alpes (France);
AvH Foundation (Germany);
ICSC (Italy); 
Severo Ochoa and Mar\'ia de Maeztu Units of Excellence, GVA, XuntaGal, GENCAT, InTalent-Inditex and Prog. ~Atracci\'on Talento CM (Spain);
SRC (Sweden);
the Leverhulme Trust, the Royal Society
 and UKRI (United Kingdom).




\clearpage
\section*{Appendices}
\appendix
\section{Tables of the relative uncertainties}
The relative uncertainties for \pt- and \y-differential prompt production cross-sections are shown in Tables~\ref{tab:uncert_pt} and \ref{tab:uncert_y}, respectively. 
The uncertainties are expressed in percent, relative to the central value in each bin. 
The total systematic uncertainty is calculated as the quadratic sum of the individual sources.
\par
\begin{sidewaystable}[ht]
    \caption{\small{Relative uncertainties (in \%) on the ratio of prompt cross-sections $\sigma_{\etac}/\sigma_{\jpsi}$ in bins of the \etac transverse momentum, \pt.
    Uncertainties on $\BR_{\etac\to\ppbar}$ and $\BR_{\jpsi\to\ppbar}$ are considered separately and given in the text.}}
    \centering\small
    \begin{tabular}{lcccccc} \hline
\multirow{2}{*}{} & \multicolumn{6}{l}{\pt [\gevc]} \\ \cmidrule{2-7}
& $5.0 - 6.5$ & $6.5 - 8.0$ & $8.0 - 10.0$ & $10.0 - 12.0$ & $12.0 - 14.0$ & $14.0 - 20.0$ \\ \hline 
Statistical uncertainty                     
& 95.6 	& 30.7 	& 14.3 	& 22.9 	& 21.8 	& 31.9 	\\ \hline
Combinatorial background                
& 4.0 	& 3.9 	& 3.8 	& 3.6 	& 3.5 	& 3.2 	\\ 
Contribution from \JpsiToPpbarPiz     
& 0.1 	& $<0.1$ 	& 0.1 	& 0.2 	& 0.3 	& 0.5 	\\ 
Resolution \pt-dependence  
& 0.1 	& $<0.1$ 	& 0.1 	& 0.3 	& 0.4 	& 0.6 	\\ 
Cross-feed                            
& $<0.1$ & 0.1 	& 0.1 	& 0.1 	& 0.2 	& 0.3 	\\ 
Total uncorrelated systematic                    
& 4.0 	& 3.9 	& 3.8 	& 3.6 	& 3.5 	& 3.4 	\\ \hline
Variation of $\Gamma_{\etac}$         
& 1.5 	& 1.1 	& 1.7 	& 1.6 	& 1.7 	& 1.8 	\\ 
Mass resolution model                 
& 3.5 	& 2.4 	& 1.6 	& 2.2 	& 1.1 	& 0.6 	\\ 
Polarisation of \jpsi                     
& 2.1 	& 1.8 	& 1.6 	& 1.3 	& 1.2 	& 0.9 	\\ 
Total correlated systematic                      
& 3.3 	& 3.1 	& 2.9 	& 2.8 	& 2.8 	& 2.6 	\\ \hline
Total systematic                            
& 5.1 	& 4.9 	& 4.8 	& 4.6 	& 4.5 	& 4.3 	\\ \hline
    \end{tabular}

    \label{tab:uncert_pt}
\end{sidewaystable}

\begin{table}[ht]
    \centering\small
    \caption{\small{Relative uncertainties (in \%) on the ratio of prompt cross-sections $\sigma_{\etac}/\sigma_{\jpsi}$ in bins of the \etac rapidity, \y.
    Uncertainties on $\BR_{\etac\to\ppbar}$ and $\BR_{\jpsi\to\ppbar}$ are considered separately and given in the text.}}
    \begin{tabular}{lcccc}\hline
\multirow{2}{*}{} & \multicolumn{4}{l}{\y} \\ \cmidrule{2-5}
 & 2.0 -- 2.5 & 2.5 -- 3.0 & 3.0 -- 3.5 & 3.5 -- 4.0 \\ \hline
Statistical uncertainty                     
& 23.3 	& 16.6 	& 22.3 	& 48.1 	\\ \hline
Combinatorial background                 
& 6.5 	& 5.4 	& 4.3 	& 3.2 	\\ 
Contribution from \JpsiToPpbarPiz     
& 0.4 	& 0.2 	& $<0.1$ 	& 0.2 	\\ 
Resolution \y-dependence  
& 1.0 	& 0.6 	& 0.3 	& $<0.1$ 	\\ 
Cross-feed                            
& 0.1 	& 0.1 	& 0.1 	& 0.1 	\\ 
Total uncorrelated systematic                    
& 6.6 	& 5.5 	& 4.3 	& 3.2 	\\ \hline
Variation of $\Gamma_{\etac}$         
& 1.0 	& 1.4 	& 1.4 	& 1.4 	\\ 
Mass resolution model                 
& 0.1 	& 1.2 	& 0.3 	& 1.7 	\\ 
Polarisation of \jpsi                       
& 2.2 	& 1.5 	& 1.2 	& 1.4 	\\ 
Total correlated systematic                      
& 3.3 	& 2.9 	& 2.8 	& 2.8 	\\ \hline
Total systematic                           
& 7.4 	& 6.2 	& 5.1 	& 4.3 	\\ \hline
    \end{tabular}

    \label{tab:uncert_y}
\end{table}
\clearpage
\section{Tables of differential production cross-sections}
\subsection{Relative differential production cross-sections}

The relative \pt- and \y-differential \etac to \jpsi prompt production cross-sections are shown in Tables~\ref{tab:relCSPT} and \ref{tab:relCSY}. 
The first uncertainty is statistical, the second is the uncorrelated systematic uncertainty, the third is the systematic uncertainty correlated among bins, and the last one is the uncertainty related to $\BR_{\JpsiToPpbar}$ and $\BR_{\EtacToPpbar}$ branching fractions. 

\begin{table}[ht]
    \centering
    \small
    \caption{\small{Relative \pt-differential \etac prompt production cross-section.
    }}
    \begin{tabular}{r@{\:$-$\:}lc}
    \hline 
    \multicolumn{2}{l}{\pt [\gevc]}  & {$d\sigma_{\etac}/d\sigma_{\jpsi}$} \\ 
    \hline 
    5.0 & 6.5   & $0.52 \pm 0.49 \pm 0.03 \pm 0.02 \pm 0.05$  \\
    6.5 & 8.0   & $1.07 \pm 0.33 \pm 0.06 \pm 0.03 \pm 0.10$  \\
    8.0 & 10.0  & $1.66 \pm 0.24 \pm 0.09 \pm 0.05 \pm 0.16$  \\
    10.0 & 12.0 & $1.43 \pm 0.33 \pm 0.11 \pm 0.04 \pm 0.14$  \\
    12.0 & 14.0 & $2.04 \pm 0.45 \pm 0.22 \pm 0.06 \pm 0.20$  \\ 
    14.0 & 20.0 & $1.89 \pm 0.60 \pm 0.24 \pm 0.05 \pm 0.18$  \\  \midrule
    5.0 & 14.0  & $1.29 \pm 0.15 \pm 0.07 \pm 0.04 \pm 0.13$  \\
    6.4 & 14.0  & $1.42 \pm 0.15 \pm 0.05 \pm 0.04 \pm 0.14$  \\
    5.0 & 20.0  & $1.32 \pm 0.14 \pm 0.08 \pm 0.04 \pm 0.13$  \\
    \hline 
    \end{tabular} 
    \label{tab:relCSPT}
\end{table} 
\par

\begin{table}[ht]
    \centering
    \small
    \caption{\small {Relative \y-differential \etac prompt production cross-section.
    }}
    \label{tab:relCSY}
    \begin{tabular}{cc}
    \hline 
    \y  & {$d\sigma_{\etac}/d\sigma_{\jpsi}$} \\ 
    \hline 
    2.0 - 2.5 & $0.96 \pm 0.22 \pm 0.07 \pm 0.03 \pm 0.09$ \\
    2.5 - 3.0 & $1.65 \pm 0.27 \pm 0.11 \pm 0.05 \pm 0.16$ \\
    3.0 - 3.5 & $1.36 \pm 0.30 \pm 0.08 \pm 0.04 \pm 0.13$ \\
    3.5 - 4.0 & $1.29 \pm 0.62 \pm 0.07 \pm 0.04 \pm 0.13$ \\
    \hline 
    \end{tabular} 
\end{table} 
\par

\subsection{Absolute differential production cross-sections}

The absolute \pt- and \y-differential \etac prompt production cross-sections are shown in Tables~\ref{tab:absCSPT} and \ref{tab:absCSY}. 
The first uncertainty is statistical, the second is the uncorrelated systematic uncertainty, the third is the systematic uncertainty correlated among bins, and the last one is the uncertainty related to $\BR_{\JpsiToPpbar}$ and $\BR_{\EtacToPpbar}$ branching fractions and the \jpsi production cross-section.
\begin{table}[ht]
    \centering
    \caption{\small{Absolute \pt-differential \etac prompt production cross-section.
    }}
    \label{tab:absCSPT}
    \centering
    \small
    \begin{tabular}{r@{\:$-$\:}lr@{\:$\pm$\:}r@{\:$\pm$\:}r@{\:$\pm$\:}r@{\:$\pm$\:}r}
    \hline 
    \multicolumn{2}{l}{\pt [\gevc]}    & \multicolumn{5}{c}{$d\sigma_{\etac}/d\pt$ $[\nb/\gevc\,]$}     \\ 
    \hline 
    5.0 &   6.5   &  277.0  & 264.8 & 16.9 & 9.0  & 29.4   \\
    6.5 &   8.0   &  227.8  & 69.8  & 12.1 & 7.0  & 24.2   \\
    8.0 &   10.0  &  134.5  & 19.2  & 7.6  & 3.9  & 14.3   \\
    10.0 &  12.0  &  42.4   & 9.7   & 3.2  & 1.2  & 4.6    \\
    12.0 &  14.0  &  23.1   & 5.0   & 2.5  & 0.6  & 2.6    \\ \hline 
    6.5 &   14.0  &  796.4  & 86.6  & 30.8 & 21.4 & 84.4   \\ 
    5.0 &   14.0  &  1772.8 & 205.4 & 97.6 & 49.0 & 187.9  \\ 
    \hline
    \end{tabular} 
\end{table} 
\par

\begin{table}[ht]
    \centering
    \small
    \caption{\small {Absolute \y-differential \etac prompt production cross-section.
    }}
    \label{tab:absCSY}
    \begin{tabular}{cr@{\:$\pm$\:}r@{\:$\pm$\:}r@{\:$\pm$\:}r@{\:$\pm$\:}r}
    \hline 
    \y  & \multicolumn{5}{c}{$d\sigma_{\etac}/d\y$ $[\nb\,]$}     \\
    \hline 
    $2.0 - 2.5$ & 1620 & 377 & 121 & 53 & 172  \\
    $2.5 - 3.0$ & 2498 & 414 & 162 & 73 & 264  \\
    $3.0 - 3.5$ & 1762 & 392 & 99 & 49 & 188 \\
    $3.5 - 4.0$ & 1291 & 621 & 69 & 37 & 139 \\
    \hline 
    \end{tabular} 
\end{table} 
\clearpage


\addcontentsline{toc}{section}{References}
\bibliographystyle{LHCb}
\bibliography{main,standard,LHCb-PAPER,LHCb-CONF,LHCb-DP,LHCb-TDR}

\newpage
\centerline
{\large\bf LHCb collaboration}
\begin
{flushleft}
\small
R.~Aaij$^{36}$\lhcborcid{0000-0003-0533-1952},
A.S.W.~Abdelmotteleb$^{55}$\lhcborcid{0000-0001-7905-0542},
C.~Abellan~Beteta$^{49}$,
F.~Abudin{\'e}n$^{55}$\lhcborcid{0000-0002-6737-3528},
T.~Ackernley$^{59}$\lhcborcid{0000-0002-5951-3498},
A. A. ~Adefisoye$^{67}$\lhcborcid{0000-0003-2448-1550},
B.~Adeva$^{45}$\lhcborcid{0000-0001-9756-3712},
M.~Adinolfi$^{53}$\lhcborcid{0000-0002-1326-1264},
P.~Adlarson$^{79}$\lhcborcid{0000-0001-6280-3851},
C.~Agapopoulou$^{13}$\lhcborcid{0000-0002-2368-0147}
% ...

{\footnotesize \it

% ...
$^{13}$Universit{\'e} Paris-Saclay, CNRS/IN2P3, IJCLab, Orsay, France\\
% ...
$^{36}$Nikhef National Institute for Subatomic Physics, Amsterdam, Netherlands\\
% ...
$^{45}$Instituto Galego de F{\'\i}sica de Altas Enerx{\'\i}as (IGFAE), Universidade de Santiago de Compostela, Santiago de Compostela, Spain\\
% ...
$^{49}$Physik-Institut, Universit{\"a}t Z{\"u}rich, Z{\"u}rich, Switzerland\\
% ...
$^{53}$H.H. Wills Physics Laboratory, University of Bristol, Bristol, United Kingdom\\
% ...
$^{55}$Department of Physics, University of Warwick, Coventry, United Kingdom\\
% ...
$^{58}$School of Physics and Astronomy, University of Glasgow, Glasgow, United Kingdom\\
$^{59}$Oliver Lodge Laboratory, University of Liverpool, Liverpool, United Kingdom\\
% ...
$^{67}$Syracuse University, Syracuse, NY, United States\\
% ...
$^{79}$Department of Physics and Astronomy, Uppsala University, Uppsala, Sweden, associated to $^{58}$\\
% ...
}
\end{flushleft}


\end{document}